\documentclass[a4paper,11pt]{article}
\pdfoutput=1 

\usepackage{jheppub} 

\usepackage[T1]{fontenc} 
\usepackage{slashed}

\newcommand{\bfm}[1]{\mbox{\boldmath$#1$}}

\newcommand{\gsim}{\;\rlap{\lower 3.5 pt \hbox{$\mathchar \sim$}} \raise 1pt
\hbox {$>$}\;}
\newcommand{\lsim}{\;\rlap{\lower 3.5 pt \hbox{$\mathchar \sim$}} \raise 1pt
\hbox {$<$}\;}

\title{\boldmath   Zero Modes of Fermions Trapped by  Giant Vortices}

\preprint{ALBERTA-THY-1-23}

\author[a]{Logan Gates} \author[] {and}
\author[a]{Alexander A. Penin}

\affiliation[a]{Department of Physics, University of Alberta,
Edmonton AB T6G 2E1, Canada
}

\emailAdd{lgates@ualberta.ca}
\emailAdd{penin@ualberta.ca}

\abstract{Zero-energy solutions of the Dirac equation for the
fermions bound to giant vortices of large winding number $n$
are studied in the  abelian Higgs and Chern-Simons Higgs
models. The case of Jackiw-Rossi  theory of the Majorana states
in topological superconductors is discussed in detail. By
expanding in inverse powers of $n$ we find an analytic result
for asymptotically  all  $n$ solutions required by the index
theorem.  In the abelian Higgs model the zero modes fill the
vortex core and reveal a {universal} structure independent of
fine details of the gauge and scalar field interactions which,
in particular, determines the general properties of the
large-$n$ superconducting cosmic strings. On the contrary, for
the Chern-Simons Higgs vortices the zero modes are localized on
the core boundary and the explicit solution is  obtained for
the supersymmetric couplings in a self-dual background.}

\begin{document}
\maketitle
\flushbottom
\section{Introduction}
\label{sec::int}

The zero-energy bound states of Dirac fermions
\cite{Jackiw:1981ee} in the field of a vortex
\cite{Abrikosov:1956sx,Nielsen:1973cs} emerge in many physical
problems ranging  from superconducting cosmic strings in
various extensions of the standard model \cite{Witten:1984eb}
to the effective Majorana states in topological superconductors
\cite{Alicea:2012ux} and Chern-Simons effective field
theory of the fractional quantum Hall effect \cite{Zhang:1989}.
These {\it zero modes} have been extensively studied in various
gauge models supporting the vortex solutions
\cite{Jackiw:1984ji,Grignani:1990iv,Lee:1992yc,Penin:1997iv}.
Their existence is often predicted by the index theorem
\cite{Weinberg:1981eu} and can be verified by a qualitative
analysis of the corresponding linear equations. Such an
approach, however, is too coarse to catch  subtle dynamical
effects which significantly affect the structure of the
solutions, while the brute-force  numerical simulations
\cite{Shore:1989,Virtanen:2008} may be insufficient to identify
their universal properties and characteristic features. At the
same time the absence of an analytic solution  of the system of
nonlinear equations for the background vortex field has
prevented finding an explicit analytic form of the zero modes
in the past. However,   the structure of the {\it giant}
vortices carrying a large winding number $n$ drastically
simplifies in the limit $n\to\infty$
\cite{Bolognesi:2005zr,Penin:2020cxj}. In some cases the vortex
equations become integrable and the analytic solution can be
found by a systematic expansion in the inverse powers of the
winding number about the asymptotic result which converges
remarkably well down to very low $n$ \cite{Penin:2021xgr}. The
resulting background field can be used to solve the
corresponding Dirac equation explicitly. This may be of
particular interest since the giant vortices are observed
experimentally in a variety of quantum condensed matter systems
\cite{Geim:1997,Grigorieva:2007,Kramer:2009,Cren:2011}
and zero-energy states can be experimentally detected
\cite{Zhang:2021}. Recently such an analysis  of  the renowned
Jackiw-Rossi model \cite{Jackiw:1981ee} has been presented in
Ref.~\cite{Gates:2022bnv}. The model  is widely used for the
description of  the zero-energy self-conjugate quasiparticle
excitations, known as Majorana zero modes, bound to the
vortices in topological superconductors or heterostructures
\cite{Chamon:2010ks,Beenakker:2013,Elliott:2014iha}. In this
paper we give a detailed account of the analysis and extend it
to the other classes of  gauge models. In Sect.~\ref{sec::2} we
discuss the abelian Higgs theory with two different types of
the fermion coupling to the scalar field: the one motivated by
the $N=2$ supersymmetric QED in $2+1$ dimensions and the one by
Jackiw and Rossi. In Sect.~\ref{sec::3} we apply our formalism
to the  $N=2$ supersymmetric Chern-Simons theory.

\section{Abelian Higgs model}
\label{sec::2}
Let us start with a brief review of the giant vortex solution
\cite{Penin:2020cxj,Penin:2021xgr}. We consider the standard
Lagrangian for the abelian Higgs (Ginzburg-Landau) model of a
scalar field of gauge charge $e$, quartic  self-coupling
$\lambda$, and vacuum expectation value $\eta$ in $2+1$
dimensions
\begin{equation}
{L_{H}}=-{1\over 4} F^{\mu\nu}F_{\mu\nu}
+\left({D^\mu \phi}\right)^\dagger D_\mu \phi
-{\lambda\over 2}\left(\left|\phi\right|^2-\eta^2\right)^2\,,
\label{eq::LagrangeH}
\end{equation}
where $D_\mu=\partial_\mu+ieA_\mu$. It is convenient to
introduce  the rescaled dimensionless quantities
\begin{equation}
e\eta r\to r, \quad \lambda/e^2\to\lambda, \quad
\phi/\eta\to\phi,  \quad   A_\mu/\eta\to A_\mu\,,
\label{eq::bosonrescale}
\end{equation}
so that in the new variables $e=\eta=1$. Vortices are static
finite-energy topologically nontrivial solutions of the
corresponding  equations of motion. The axially symmetric
solutions  of winding number $n$  in polar coordinates can be
written as follows $\phi(r,\theta)=f(r)e^{in\theta}$,
$A_\theta=-n a(r)$, $A_r=0$, $f(0)=a(0)=0$ and
$f(\infty)=a(\infty)=1$. For a given winding number the
solution carries $n$ quanta of magnetic flux $\Phi= -\int
F_{12}{\rm d}^2{\bfm r}=2\pi n$. In the limit $n\to\infty$ the
vortices evolve into the {\it thin-wall flux tubes} of energy
$T=2\pi\sqrt{\lambda}n\eta^2$ with the nonlinear dynamics
confined to a finite-depth boundary layer wrapping the vortex
core of radius $r_n=\sqrt{2n}/\lambda^{1/4}$.  Outside the
boundary layer the dynamics of the gauge and scalar fields
linearizes up to the corrections exponentially suppressed for
large $n$. In the vortex core  $r<r_n$ it is determined solely
by the gauge interaction giving the universal solution
\begin{equation}
\begin{split}
&f(r)=F\exp\left[{n\over 2}\left(\ln\left({r^2\over r^2_n}\right)
- {r^2\over r_n^2}+1\right)\right],\\
&a(r)={r^2\over r_n^2}\,,
\label{eq::coresolH}
\end{split}
\end{equation}
where $F$ is an integration constant and $a(r)$ corresponds to
a homogeneous magnetic field. At $r>r_n$  the scalar and gauge
fields exponentially approach their vacuum values
\begin{equation}
\begin{split}
& f(r) \sim 1+ {\nu\over 2\pi}K_0(\sqrt{2\lambda}r)+\ldots\,, \\
&a(r) \sim 1+ {\mu\over 2\pi n}{\sqrt{2}r}K_1(\sqrt{2}r)+\ldots\,,
\label{eq::tailsolH}
\end{split}
\end{equation}
where $K_m(z)$ is the $m$th modified Bessel function, and the
vortex scalar charge $\nu$ and magnetic dipole moment $\mu$
read
\begin{equation}
\begin{split}
&\nu\sim -e^{2\sqrt{n}\,\lambda^{1/4}+\ldots} \,,\qquad
\mu\sim -e^{2\sqrt{n}/\lambda^{1/4}+\ldots}\,.
\label{eq::chargelam}
\end{split}
\end{equation}
Inside the boundary layer the asymptotic vortex solution does
not depend on the winding number and gets the corrections in
powers of $1/\sqrt{n}$ \cite{Penin:2021xgr}.  Throughout
the paper we consistently use the universal aspects of the
leading order result and neglect the model-dependent
corrections. The   system of the second
order non-linear boundary layer equations  becomes
integrable in the  limits of large, critical, and small scalar
self-coupling. For the critical coupling  $\lambda=1$
corresponding to the $N=2$ supersymmetric QED
\cite{DiVecchia:1977nxl} the order of the equations reduces
\cite{Bogomolny:1975de} and  the {\it self-dual} solution can
be found analytically.\footnote{For $N=2$ supersymmetric models
with nonvanishing superpotential and flat directions of the
scalar potential \cite{Witten:1993yc} the structure of the
vortex solution is significantly different \cite{Penin:1996si}
and is not discussed in the present paper.} For the functions
\begin{equation}
\begin{split}
&w(x)=\ln f(r_n+x)\,,\\
&\gamma(x)={n}\left(a(r_n+x)-1\right)/r_n\,,
\label{eq::omgamdef}
\end{split}
\end{equation}
it reads
\begin{equation}
\begin{split}
& \int^{w(x)}_{w_0} {{\rm d}w \over (e^{2 w}-2w-1)^{1/2}}=x  \,,\\
& \gamma(x) =  -(e^{2 w(x)}-2w(x)-1)^{1\over 2}\,,
\label{eq::boundsolH}
\end{split}
\end{equation}
where
\begin{equation}
w_0=-0.2997174398\ldots\,,
\label{eq::w0}
\end{equation}
with the following  asymptotic behavior at $x\to\infty$
\begin{equation}
\begin{split}
& w(-x) \sim -{x^2\over 2}-{1\over 2}\,,\\
&w(x) \sim w_0\exp\left[\int_{w_0}^0\left({\sqrt{2}\over (e^{2w}
-2w-1)^{1/2}}+{1\over w}\right){\rm d}w\right]e^{-\sqrt{2}x}  \,.
\label{eq::wasym}
\end{split}
\end{equation}
In the limit $1\ll\lambda\ll n^2$ it becomes
\begin{equation}
\begin{split}
&f(r_n+x) =\left\{
\begin{array}{ll}
\left(1-2\,{\rm sech}^2
\left(\sqrt{2}(x+1-x_0)\right)\right)^{1\over 2} \,,  & x\ge -1\,, \\
&\\[-3mm]
0\,, \quad & x<-1 \,, \\
\end{array}
\right.
\\[1mm]
& \gamma(x) = -{\sqrt{2\lambda}}\,{\rm sech}
\left(\sqrt{2}(x+1-x_0)\right)\,,
\label{eq::largelamsol}
\end{split}
\end{equation}
where $x_0=-{\rm arcsinh}(1)/\sqrt{2}$. The solution in the
opposite limit  $1/n^2\ll\lambda\ll 1$ is discussed in detail
in \cite{Penin:2021xgr}. As we discuss in the next section in
the abelian Higgs model only a finite number {\it i.e.} a
vanishing fraction of the total number of zero modes  are
affected by the nonlinear dynamics of the boundary layer, hence
the absence of the general solution is not crucial.

Yukawa interaction of the massless Dirac fermions to the scalar
field is mandatory for the existence of the normalizable zero
modes in the vortex background \cite{Jackiw:1981ee}. Below we
consider two basic types of the Yukawa interaction: the fermion
number preserving coupling inspired by the $N=2$ supersymmetry
and the fermion number violating coupling describing the
quasiparticle interaction to the Cooper pair condensate in
topological superconductors.

\subsection{Supersymmetry motivated  fermion coupling}
\label{sec::2.1}
\begin{figure}[t]
\begin{center}
\includegraphics[width=10cm]{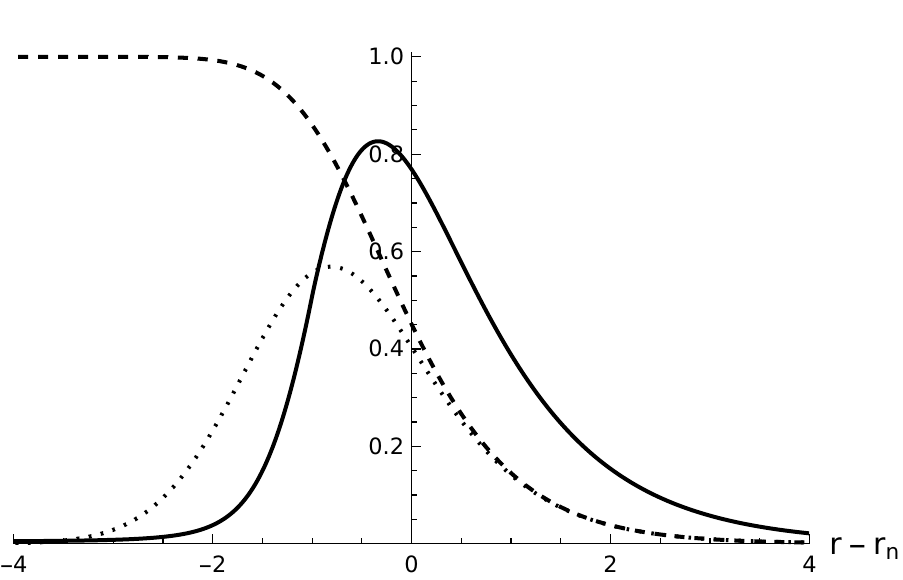}
\end{center}
\caption{\label{fig::1} The supersymmetric   zero modes
$\sqrt{n}\psi^+_{n-1}(r)$ (dotted line) and $-\sqrt{
n}\chi^-_{0}(r)$ (dashed line) of the abelian Higgs  model,
Eq.~(\ref{eq::solSH}). The normalized boundary layer zero mode
$\sqrt{r_n}\psi^+_{n/2}(r)$ (solid line) of  the Jackiw-Rossi model for
$\lambda=16$ and  $g=1$, Eq.~(\ref{eq::zeromJRllam}).}
\end{figure}

The fermion number preserving  Yukawa interaction to the
charged scalar field requires a pair of charged and neutral
fermions $\psi$ and $\chi$ with the Lagrangian density
\cite{DiVecchia:1977nxl}
\begin{equation}
{L_{F}}=i\bar{\psi}\slashed{D}\psi+i\bar{\chi}
\slashed{\partial}\chi+i\sqrt{2}g
\left(\bar{\psi}\chi\phi-\bar{\chi}\psi\phi^*\right)\,,
\label{eq::LagrangeF1}
\end{equation}
where the three-dimensional Dirac matrices reduce to the Pauli
matrices $\gamma^\mu=(\sigma_3,i\sigma_2,-i\sigma_1)$, and the
gauge charges of the fermion and the scalar fields are equal to
$e$. As before we introduce the rescaled dimensionless
quantities
\begin{equation}
g/e\to g, \quad \psi/\sqrt{e\eta}\to \psi, \quad \chi/\sqrt{e\eta}\to \chi\,.
\label{eq::fermresc}
\end{equation}
Then for $\lambda=g=1$ the Lagrangian
Eqs.~(\ref{eq::LagrangeH},\,\ref{eq::LagrangeF1}) is invariant
under the $N=2$ supersymmetry  transformations with $\psi$
($\chi$)  being a part of the chiral (vector) supermultiplet.
For  arbitrary $g$ the static zero-energy modes are determined
by the following equations for the spinor components
\begin{equation}
\begin{split}
& D_+\psi^+-\sqrt{2}g\phi\chi^- =0\,, \\
& \partial_-\chi^- -\sqrt{2}g\phi^*\psi^+ =0\,,
\label{eq::Diraceq1a}
\end{split}
\end{equation}
\begin{equation}
\begin{split}
& D_-\psi^-+\sqrt{2}g\phi\chi^+ =0\,, \\
& \partial_+\chi^+ +\sqrt{2}g\phi^*\psi^- =0\,,
\label{eq::Diraceq1b}
\end{split}
\end{equation}
where the chiral derivative is
\begin{equation}
D_\pm=D_1\pm iD_2=e^{\pm i\theta}
\left[\partial_r\pm \left({i}
\partial_\theta+na(r)\right)/r\right]\,.
\label{eq::Dpm}
\end{equation}
Eqs.~(\ref{eq::Diraceq1a}) and (\ref{eq::Diraceq1b}) have
normalizable solutions for positive and negative values of $n$,
respectively. Below we consider the case $n>0$. Then by
the decomposition
\begin{equation}
\begin{split}
&\psi^+=\sum_{l=0}^{n-1}{e^{il\theta}\over \sqrt{2\pi}}\psi^+_l \,, \\
&\chi^-=\sum_{l=-n+1}^{0}{e^{il\theta}\over \sqrt{2\pi}}\chi^-_{l} \,,
\label{eq::pwexpH}
\end{split}
\end{equation}
Eq.~(\ref{eq::Diraceq1a}) can by transformed into the equation
for the partial waves
\begin{equation}
\begin{split}
&\left({d\over dr}-{l\over r}+{n a\over r}\right)\psi_l^+
-\sqrt{2}gf\chi^-_{-n+l+1}=0\,,\\
&\left({d\over dr}-{n-l-1\over r}\right)\chi^-_{-n+l+1}
-\sqrt{2}gf\psi^+_{l}=0\,.
\end{split}
\label{eq::pweqH}
\end{equation}
Outside the core for $r>r_n$ the vortex fields approach the
vacuum values $a(r),~f(r)\approx 1$  and Eq.~(\ref{eq::pweqH})
describes free fermions of the mass $\sqrt{2}|g|$. Its
normalizable solution exponentially decays as
$\psi_l^+,~\chi^-_{-n+l+1}\propto e^{-\sqrt{2}|g|r}$ so that
the zero modes are localized in the vortex core or on its
boundary. Inside the core  the function $f$ is exponentially
suppressed at large $n$. Neglecting the Yukawa term  in
Eq.~(\ref{eq::pweqH}) we find $\chi^-_{-n+l+1}\propto
r^{n-l-1}$  exponentially suppressed too, except the partial
waves with  $n-l={\cal O}(1)$ which will be discussed
separately. Thus, the neutral component of the zero mode
asymptotically vanishes and the equation for the remaining
charged component at $r<r_n$ becomes
\begin{equation}
\begin{split}
&\left({d\over dr}-{l\over r}
+{\lambda^{1/2}\over 2}r\right)\psi_l^+=0\,
\end{split}
\label{eq::pweqpsi}
\end{equation}
with the solution
\begin{equation}
\begin{split}
&\psi^+_l(r)\sim N_{l,2}\,r^le^{-\lambda^{1/2}r^2/4} \,, \\
\label{eq::zerom1}
\end{split}
\end{equation}
where
\begin{equation}
N_{l,c}=\left({2\lambda^{l+1\over 2}\over c^{l+1} l!}\right)^{1\over 2}
\label{eq::NL}
\end{equation}
is the normalization factor. Eq.~(\ref{eq::zerom1}) describes
the approximately Gaussian peaks of width
$\sigma=1/\lambda^{1/4}$  centered at ${r}_l$
\begin{equation}
\psi^+_l(r)\approx N'_{l,2}\,e^{-\lambda^{1/2}(r-r_l)^2/2}\,,
\label{eq::zeromG}
\end{equation}
where the normalization factor is
\begin{equation}
N'_{l,c}=\left({\lambda\over c^2\pi^2 l}\right)^{1\over 4}\,.
\label{eq::NG}
\end{equation}
This solution has a clear physical interpretation - the charged
fermion forms the lowest Landau level states in the homogeneous
magnetic field of the giant vortex core with each  state
encircling  an integer number of the  flux quanta. However, a
few states with $l\approx n$ are  localized near the core
boundary where the  nonlinear effects become crucial and the
approximation breaks down. In this case the closed form
analytic solution is not available but for $n\to\infty$ the
fraction of the zero modes localized at the boundary vanishes.
Moreover,  for  $\lambda=g=1$
the $l=n-1$ solution can be obtained by supersymmetry
transformation of the vortex fields \cite{Penin:1996si}
\begin{equation}
\begin{split}
&\psi^+=D_-\phi \,, \\
&\chi^-= {1\over \sqrt{2}}\left[F_{12}
+\left(\left|\phi\right|^2-1\right)\right]\,.
\label{eq::zeromS}
\end{split}
\end{equation}
By substituting  the critical vortex solution of  the previous
section in the boundary layer region $r-r_n={\cal O}(1)$ we get
the normalized supersymmetric  zero mode
\begin{equation}
\begin{split}
&\psi^+_{n-1}={e^w(e^{2w}-1-2w)^{1\over 2}\over \sqrt{n/2}}\,, \\
&\chi^-_0= {e^{2w}-1\over \sqrt{n}}\,,
\label{eq::solSH}
\end{split}
\end{equation}
where $w(r-r_n)$ is given by Eq.~(\ref{eq::boundsolH}). In
Eq.~(\ref{eq::solSH}) the result for the neutral component
$\chi^-_0$ matches the nonvanishing constant core solution,
and the  shape of the charged component is a non-Gaussian
peak at $r=r_n$ which decays as $e^{-(r-r_n)^2/2}$ for $r<r_n$
and only as $e^{-\sqrt{2}(r-r_n)}$ for $r>r_n$. This gives a
qualitative picture of the boundary layer zero modes for
nonsupersymmetric  couplings $g,~\lambda={\cal O}(1)$ as well.
The shape of the supersymmetric solution is shown in
Fig.~\ref{fig::1}.

\subsection{Jackiw-Rossi fermion coupling}
\label{sec::2.2}

The Lagrangian density of the Jackiw-Rossi model reads
\cite{Jackiw:1981ee}
\begin{equation}
{L_{F}}=i\bar{\psi}\slashed{D}\psi+{g\over 2}
\left(\bar{\psi}\psi^c\phi+\bar{\psi}^c\psi\phi^*\right)\,,
\label{eq::LagrangeF2}
\end{equation}
where $\psi^c=-i\sigma_1\psi^*$ is the charge conjugate spinor
and the gauge invariance requires the charge of the fermion
to be a half of the scalar field charge. The zero-mode equation
now takes the form
\begin{equation}
D_\pm\psi^\pm+g\phi{\psi^*}^\pm =0\,.
\label{eq::Diraceq2}
\end{equation}
For positive $n$ the negative chirality equation
does not have a normalizable solution. The $n$ zero modes of
positive chirality do not have definite angular momentum
and are given by  the linear combinations of the partial waves
\begin{equation}
\begin{split}
&\xi^+_l={1\over 2\sqrt{\pi}}\left(e^{il\theta}\psi^+_l +e^{i(n-1-l)\theta}\psi^+_{n-1-l}\right)\,,\\
&\eta^+_l={i\over  2\sqrt{\pi}}\left(e^{il\theta}\psi^+_l -e^{i(n-1-l)\theta}\psi^+_{n-1-l}\right)\,,
\label{eq::zmdec2}
\end{split}
\end{equation}
where $0\le l\le n/2-1$ for even $n$ and $0\le l\le (n-1)/2$,
$\eta^+_{(n-1)/2}=0$ for odd $n$. The equations for the
partial waves read
\begin{equation}
\begin{split}
& \left({d\over dr}-{l\over r}+{na\over 2r}\right)
\psi_l^++gf\psi_{n-1-l}^+=0
\,, \\
&\left({d\over dr}-{n-l-1\over r}+{na\over 2r}\right)
\psi_{n-1-l}^++gf\psi_{l}^+=0\,,
\label{eq::pweqJR}
\end{split}
\end{equation}
and differ from Eq.~(\ref{eq::pweqH}) by the gauge field term.
After identification of $\psi_{l}$ and $\psi_{n-1-l}$ with the
components of the Nambu spinor,  and of $f$ with the pair
potential the above system reproduces the Bogoliubov-de-Gennes
equations for the  Majorana vortex zero modes of the effective
Dirac fermion at zero chemical potential in the condensed
matter systems (see {\it e.g.} \cite{Chamon:2010ks}). It is
convenient to decouple the  gauge field  by a field
redefinition $\psi_l^+(r)=u_l(r)G(r)$, where the gauge factor
\begin{equation}
G(r)=\exp\left(-{n\over 2}\int_0^r {a(r')\over r'}{\rm d}r'\right)
\label{eq::Gdef}
\end{equation}
at large $n$ evaluates to
\begin{equation}
G(r)\sim\left\{
\begin{array}{ll}
 e^{-\lambda^{1\over 2}r^2/8}\,,  & r<r_n \,, \\
 (r_n/r)^{n\over 2} e^{-\lambda^{1\over 2}r_n^2/8}\,, & r>r_n \,. \\
\end{array}
\right.
\label{eq::Gres}
\end{equation}
Then the system Eq.~(\ref{eq::pweqJR}) can be transformed
into the second order equation
\begin{equation}
\left[{d^2\over dr^2}-\left({n-1\over r}+{f'\over f}\right){d\over dr}
+{l\over r}\left({n-l\over r}+{f'\over f}\right)-g^2f^2\right]u_l
=0\,,
\label{eq::uleq}
\end{equation}
where $f'=df/dr$. Outside the core $f\approx 1$, $f'\approx 0$,
and its normalizable solution is $u_l(r)=r^{n/2}K_{\mu}(|g| r)$,
$\mu=\sqrt{n^2/4-l(n-l)}$, which gives
\begin{equation}
\psi^+_l(r)\propto K_{\mu}(|g|r)\,.
\label{eq::ultail}
\end{equation}
Though inside the vortex core the scalar field is exponentially
suppressed,  it is a singular perturbation since the order of
the system Eq.~(\ref{eq::pweqJR}) for vanishing $f$ is reduced.
It cannot be in general neglected to get two  solutions regular
at $r=0$ which are required to match  Eq.~(\ref{eq::ultail})
for $r\to\infty$. For the model discussed in the previous
section the regular solution missing for vanishing $f$ is
exponentially suppressed. However, in the case under
consideration both solutions are relevant. Indeed for $r<r_n$
the logarithmic derivative term $f'/f={n/
r}\left(1-{r^2/r_n^2}\right)$ in Eq.~(\ref{eq::uleq}) is not
suppressed and the two regular  solutions are
\begin{equation}
\begin{split}
&u_l^{(1)}(r)=r^l\,, \\
&u_l^{(2)}(r)=r^{2n-l}{\rm E}_{\nu}\left(n r^2/2 r_n^2\right)\,,
\label{eq::ulsol}
\end{split}
\end{equation}
where ${\rm E}_{\nu}(z)$ is the $\nu$th exponential integral
with $\nu = 1+l-n$. The behavior of the second solution at
large $n$ is quite peculiar. For $l<n/2$ it reduces to
$u_l^{(2)}(r)\sim r^l$ {\it i.e.} the two solutions are
degenerate up to the exponentially suppressed terms. For  $l>
n/2$, however,  it transforms into $u_l^{(2)}(r)\sim
r^{2n-l}e^{-n r^2/2r_n^2}$. For such $l$ the first solution is
exponentially suppressed and is the only solution which gives
an unsuppressed contribution to $\psi_l^+(r)$ is $u_l^{(2)}$.
For the partial waves in the large-$n$ limit we finally get
\begin{equation}
\psi^+_l(r)\sim \left\{
\begin{array}{ll}N_{l,4}\,
r^l e^{-\lambda^{1\over 2} r^2/8} \,,  & l<n/2 \,,\\
N_{2n-l,{4\over 3}}\,
r^{2n-l}e^{-3\lambda^{1\over 2} r^2/8}\,, \quad & l>n/2 \,. \\
\end{array}
\right.
\label{eq::zeromJR}
\end{equation}
Eq.~(\ref{eq::zeromJR}) describes two groups of approximately
Gaussian peaks
\begin{equation}
\psi^+_l(r)\approx\left\{
\begin{array}{ll}
N'_{l,4}\,
e^{-\lambda^{1\over 2}(r-\bar{r}_l)^2/4}\,,  & l<n/2\,, \\
N'_{2n-l,{4\over 3}}\,
e^{-3\lambda^{1\over 2}(r-\bar{r}'_l)^2/4}\,, \quad & l>n/2\,.  \\
\end{array}
\right.
\label{eq::zeromJRG}
\end{equation}
For $l<n/2$ the peaks of the width
$\sigma=\sqrt{2}/\lambda^{1/4}$ are centered at
$\bar{r}_l=\sqrt{2l/n} \,r_n$, while for $l>n/2$ the peaks have
the width $\sigma'=\sigma/\sqrt{3}$ and  are centered at
$\bar{r}'_l=\sqrt{(2/3)(2-l/n)}\,r_n$.
The physical origin of the second group of states is as
follows. Due to $e/2$ charge each of the lowest Landau level states
now encircles  an even number of the  flux quanta. Hence, only
about $n/2$   of the Landau states fit into the vortex core and
are not affected by the vortex scalar field. For larger $l$ the
effect of the scalar field on the localization of the states
increases, as it can be seen in Eq.~(\ref{eq::uleq}),
and for $l\approx n$ it exceeds the effect of the
magnetic field, resulting in a set of {\it warped} lowest Landau level
states.  Remarkably such a significant effect is achieved in
the region where the scalar field is exponentially small, which
can be attributed to the  singular character of the $f\to 0$
limit of the zero-mode equations discussed above. The solutions
with $l\approx n/2$ are localized inside the boundary layer
where the nonlinear effects are essential and an explicit
analytical solution is not available. At the same time for
known functions $f(r)$ and $a(r)$  the solution in the region
$r-r_n={\cal O}(1)$ is given by
\begin{equation}
\psi^+_{n/2}(r)\propto r^{n\over 2}
\exp\left[-\int^r\left({n a(r')\over 2r'}
+|g|f(r')\right){\rm d}r'\right]\,,
\label{eq::zeromJRbl}
\end{equation}
where  $n$ is assumed to be even. Note that the asymptotic
solution Eq.~(\ref{eq::zeromJRbl}) in the limit $n\to\infty$
does not depend on $n$.  For example, in the integrable case
$\lambda\gg 1$ we get
\begin{eqnarray}
\psi^+_{n/2}(r_{n}+x)&\propto&
\exp\left\{\int_0^{x+1}\bigg[\sqrt{\lambda/2}\,\,
{\rm sech}\left( \sqrt{2}(x'-x_0)\right)\right.
\nonumber\\
&-&\left.\left.|g|\left(1-2\,{\rm sech}^2
\left(\sqrt{2}(x'-x_0)\right)\right)^{1\over 2}\right]
{\rm d}x'\right\}\,.
\label{eq::zeromJRllam}
\end{eqnarray}
The shape of the solution Eq.~(\ref{eq::zeromJRllam}) for some
typical values of the coupling constants is shown in
Fig.~\ref{fig::1}. At the same time, the radial density of the
zero modes
$\rho(r)=\sum_{l=0}^{n-1}\left(\psi^+_l(r)\right)^2$, which is
the primary object in the condensed matter applications, for
large $n$ is not sensitive to the nonuniversal contribution of
the  $l\approx n/2$ states and in the $n\to\infty$  limit it
converges to a piecewise constant function
\begin{equation}
\rho(r)\sim \sqrt{\lambda}\left\{
\begin{array}{ll}
{1/ 2}\,,  & r/r_n<\sqrt{2/ 3}\,, \\
&\\[-2mm]
2\,, \quad & \sqrt{2/3}<r/r_n<1 \,, \\
\end{array}
\right.
\label{eq::rho}
\end{equation}
with   $\int_0^\infty \rho(r) r {\rm d}r = n$.

\section{Supersymmetric Chern-Simons  model}
\label{sec::3}

We consider the $N=2$ supersymmetric   model with the Lagrange
density \cite{Lee:1990it}
\begin{eqnarray}
{L_{CS}}&=&{\kappa\over 4}
\epsilon^{\mu\nu\lambda}F_{\mu\nu}A_{\lambda}
+\left({D^\mu \phi}\right)^\dagger D_\mu \phi
-{e^4\over \kappa^2}\left|\phi\right|^2
\left(\left|\phi\right|^2-\eta^2\right)^2
\nonumber \\
&&+i\bar{\psi}\slashed{D}\psi+{e^2\over \kappa}
\left(3\left|\phi\right|^2-\eta^2\right)\bar{\psi}\psi\,,
\label{eq::LagrangeCS}
\end{eqnarray}
where the  field equations can be solved for the
nondynamical scalar potential
\begin{equation}
 A_0=-{\kappa\over 2e^2\eta^2}{F_{12}\over |\phi|^2}\,.
\label{eq::A0sol}
\end{equation}
This theory supports the self-dual vortex solutions
\cite{Jackiw:1990aw} and the corresponding first-order
equations read
\begin{equation}
\begin{split}
& D_+\phi =0\,, \\
& F_{12}-{2e^3\over \kappa^2}\left|\phi\right|^2
\left(\left|\phi\right|^2-\eta^2\right)=0\,.
\label{eq::BogomolnyCS}
\end{split}
\end{equation}
For the axially symmetric vortex configuration with
the winding number $n$ they reduce to
\begin{equation}
\begin{split}
& {df\over dr}-{n\over r}\left(1-a\right)f  = 0 \,, \\
& {da\over dr}+{r\over n}{2\over \kappa^2}f^2(f^2-1) = 0\,,
\label{eq::eqCS}
\end{split}
\end{equation}
where the rescaled dimensionless parameter
$\eta\kappa/e\to\kappa$ is introduced along with the
dimensionless fields and couplings  defined in the previous
sections. To get the large-$n$ asymptotic solution of
Eq.~(\ref{eq::eqCS}) we follow the approach
\cite{Penin:2020cxj}. In contrast to the Abrikosov giant
vortices the boundary layer develops at $r_n=n\kappa$. Inside
the core the  dynamics of  the scalar field  linearizes and
Eq.~(\ref{eq::eqCS}) with the exponential accuracy can be
approximated as follows
\begin{equation}
\begin{split}
& {df\over dr}-{n\over r}f  = 0 \,, \\
& {da\over dr}-{r\over n}{2\over \kappa^2}f^2 = 0\,,
\label{eq::coreeqCS}
\end{split}
\end{equation}
where the leading  nonvanishing term is left in the second
line. The solution of  Eq.~(\ref{eq::coreeqCS})  is
\begin{equation}
\begin{split}
&f(r) \sim \left({r/ r_n}\right)^n\,, \\
&a(r) \sim \left({r/r_n}\right)^{2n}\,,
\label{eq::corsolCS}
\end{split}
\end{equation}
where the integration constant is found by matching to the
boundary layer solution. To get the asymptotic result inside
the boundary layer  for $r-r_n={\cal O}(1)$ one neglects the
difference between $r$ and $r_n$ in the coefficients of
Eq.~(\ref{eq::eqCS}), which transforms into
\begin{equation}
\begin{split}
& w''-2e^{4w}+2e^{2w}  = 0 \,, \\
& a(r_n+x) = 1-w'(x/\kappa)\,,
\label{eq::boundeqCS}
\end{split}
\end{equation}
where $w(x/\kappa)=\ln(f(r_n+x))$. The first line of
Eq.~(\ref{eq::boundeqCS}) has a first integral $I=
{w'}^2-e^{4w}+2e^{2w}=1$ and can be easily integrated with
proper boundary conditions at $x=\pm \infty$ yielding
\begin{equation}
\begin{split}
& f(r_n+x)= \left(1+ e^{-2x/\kappa}\right)^{-1/2}\,, \\
&a(r_n+x)= \left(1+ e^{-2x/\kappa}\right)^{-1}\,,
\label{eq::boundsolCS}
\end{split}
\end{equation}
which coincides with  Eq.~(\ref{eq::corsolCS}) in the matching
region $1\ll |x|\ll r_n$ for negative $x$, where both
approximations are valid. Eq.~(\ref{eq::boundsolCS}) agrees
with the result of Ref.~\cite{Bolognesi:2007ez}. The ${\cal
O}(1/n)$ corrections to  this result can be obtained within the
approach \cite{Penin:2021xgr} but are not discussed here.
Outside the core the solution of the linearized equations is
similar to Eq.~(\ref{eq::tailsolH})
\begin{equation}
\begin{split}
&f(r) \sim 1+ {\nu\over 2\pi}K_0(2 r/\kappa)\,, \\
&a(r) \sim 1+ {\mu\over 2\pi}{2 r\over \kappa n}K_1(2 r/\kappa)\,.
\label{eq::tailsolCS}
\end{split}
\end{equation}
The values of the vortex magnetic moment and scalar charge are
obtained by matching Eq.~(\ref{eq::tailsolCS}) to
Eq.~(\ref{eq::boundsolCS}) in the second matching region $1\ll
x\ll r_n$ with the result
\begin{equation}
\mu=\nu=-2\sqrt{\pi}e^{2n+\ln(n)/2}\,.
\label{eq::munuCS}
\end{equation}
Besides the  core radius, the main difference between the giant
vortices in the abelian Higgs and Chern-Simons Higgs models is
in the magnetic flux distribution. In the latter the magnetic
flux is confined to the boundary layer rather than inside the
vortex core. This results in a qualitatively different
structure and localization of the fermion zero modes as they
cannot be trapped inside the core by the magnetic field. The
zero-mode equations for the Lagrangian
Eq.~(\ref{eq::LagrangeCS}) are
\begin{equation}
\begin{split}
& D_+\psi^++i{2\over \kappa}|\phi|^2\psi^- =0\,, \\
& D_-\psi^-+i{2\over \kappa}\left(1-2|\phi|^2\right)\psi^+ =0\,.
\label{eq::DiraceqCS}
\end{split}
\end{equation}
In terms of the partial waves
\begin{equation}
\begin{split}
&\psi^+=\sum_{l=0}^{n-1}{e^{il\theta}\over \sqrt{2\pi}}\psi^+_l \,, \\
& \psi^-=\sum_{l=1}^{n}{e^{il\theta}\over \sqrt{2\pi}}\psi^-_l\,,
\label{eq::pwexpCS}
\end{split}
\end{equation}
they read
\begin{equation}
\begin{split}
& \left({d\over dr}-{l\over r}+{n a\over r}\right)\psi^+_l
+i{2\over \kappa}|\phi|^2\psi^-_{l+1} =0\,, \\
& \left({d\over dr}+{l+1\over r}-{n a\over r}\right)\psi^-_{l+1}
+i{2\over \kappa}\left(1-2|\phi|^2\right)\psi^+_l =0\,.
\label{eq::pweqCS}
\end{split}
\end{equation}
To get the large-$n$ asymptotic result for  the zero modes
localized inside the boundary layer we can approximate $r$ by
$r_n$ in the coefficients of  Eq.~(\ref{eq::pweqCS}), which
converts into the system
\begin{equation}
\begin{split}
&\kappa^2{\psi_l^+}''-2\kappa\left(1-f^2\right){\psi_l^+}'
+\left(1-\alpha^2+4f^2-9f^4\right)\psi_l^+=0\,,\\
&\psi^-_{l+1}=-i\psi_l^+\,,
\label{eq::psileqCS}
\end{split}
\end{equation}
where $\alpha=1-l/n$, and we use the relations $\kappa
f'/f=(1-f^2)$, $a=f^2$ following from the boundary layer
equations Eq.~(\ref{eq::boundeqCS}). In Eq.~(\ref{eq::psileqCS})
we consider $\alpha\ne 1$ and neglect the terms suppressed by
$1/l$ and $1/n$ in the $n\to\infty$ limit, where $0\le\alpha<
1$ becomes a continuous parameter. For a finite number of the
partial waves with $l/n\approx 0$, the zero modes extend beyond
the boundary layer and the above  approximation breaks down.
The fraction of such solutions, however, vanishes when  $n\to
\infty$.  After a change of the variables
\begin{equation}
\begin{split}
&{\psi_l^+}= N(\alpha){z^{1/2}\over(1+z)^{3/2}}g(z)\,,
\label{eq::zmdec}
\end{split}
\end{equation}
where $z=e^{2(r-r_n)/\kappa}$ and $N(\alpha)$ is the
normalization factor, we get the following equation
on $g(z)$
\begin{equation}
{d^2 g\over dz^2}+{1-z\over z(1+z)}{d g\over dz}
-{\alpha^2\over 4z^2}g=0 \,.
\label{eq::geq}
\end{equation}
\begin{figure}[t]
\begin{center}
\includegraphics[width=10cm]{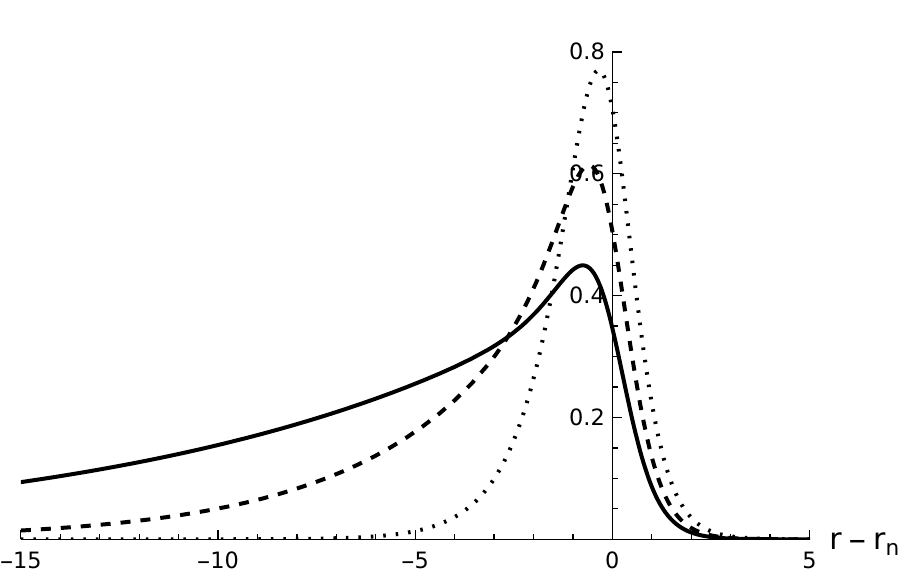}
\end{center}
\caption{\label{fig::2} The normalized   zero modes
$\sqrt{2 r_n}\psi^+_{l}(r)$  of the Chern-Simons  model,
Eq.~(\ref{eq::zmdec}), for $\kappa=1$ and $\alpha=0$ (dotted
line), $\alpha=0.75$ (dashed line), $\alpha=0.9$ (solid
line).}
\end{figure}
Its normalizable solution is given by a linear combination
of the hypergeometric  functions
\begin{eqnarray}
g(z)&=&{1\over 2\sqrt{1-\alpha}}
\Bigg[z^{\alpha/2}{}_2F_1\left(-1-(\sqrt{4+\alpha^2}-\alpha)/2,
-1+(\sqrt{4+\alpha^2}+\alpha)/2;1+\alpha;-z\right)
\nonumber\\
&-& {D(\alpha)\over D(-\alpha)}z^{-\alpha/2}
{}_2F_1\left(-1-(\sqrt{4+\alpha^2}+\alpha)/2,
-1+(\sqrt{4+\alpha^2}-\alpha)/2;1-\alpha;-z\right)
\Bigg]\,,\nonumber\\
\label{eq::gsol}
\end{eqnarray}
where
\begin{equation}
D(\alpha)=\Gamma\left(1+\alpha\right)
\Gamma\left(-1+(\sqrt{4+\alpha^2}-\alpha)/2\right)
\Gamma\left(2+(\sqrt{4+\alpha^2}-\alpha)/2\right)\,.
\label{eq::Dal}
\end{equation}
The normalization factor cannot be computed analytically for
general $\alpha$. We have found that it monotonically increases
with increasing $\alpha$ from $N(0)={\sqrt{2}/ (\kappa
\sqrt{n})}$ to
$\hspace{1mm}\lim\hspace{-7mm}\raisebox{-2mm}{\scriptsize$\alpha\to
1$}N(\alpha)= \sqrt{2}N(0)$. The shape of the solution
Eq.~(\ref{eq::zmdec}) for some typical values of $\kappa$ and
$\alpha$ is shown in Fig.~\ref{fig::2}. Let us discuss the
general properties of the solution. For  $\alpha=0$ it reduces
to  $g(z)=1$  and up to normalization describes the  zero modes
\begin{equation}
\begin{split}
&\psi^+_{n-1}=D_-\phi \,, \\
&\psi^-_{n}= i{2\over \kappa}\left(\left|\phi\right|^2-1\right)\phi\,,
\label{eq::zeromSCS}
\end{split}
\end{equation}
resulting from the supersymmetry transformation of the vortex
solution \cite{Lee:1992yc}. For $\alpha\ne 0$ the asymptotic
behavior of the solution reads
\begin{equation}
\begin{split}
& g(z)\sim C_\infty(\alpha)
z^{1-\sqrt{1+(\alpha/2)^2}}\,, \quad z\to\infty\,, \\
& g(z)\sim -{1\over 2\sqrt{1-\alpha}}{D(\alpha)
\over D(-\alpha)}{z^{-\alpha/2}}\,, \quad z\to 0\,,\\
\label{eq::gasym}
\end{split}
\end{equation}
where the constant $C_\infty(\alpha)$ has a lengthy analytic
expression which is not required for our analysis.
As it follows from  Eqs.~(\ref{eq::zmdec},\,\ref{eq::gasym})
for $1-\alpha={\cal O}(1/n)$, which corresponds to $l={\cal
O}(1)$, the zero mode solution extends to $x\approx -r_n$, {\it
i.e.} outside the region where the boundary layer approximation
can be used. For such partial waves the result
Eq.~(\ref{eq::zmdec}) is not valid as it was pointed out above.

\section{Summary}
\label{sec::sum}
Thus, we have studied the zero-energy solutions of the Dirac
equation in the giant vortex background in the limit of large
winding number $n$. For the critical scalar coupling
corresponding to the models with $N=2$ extended supersymmetry
the explicit asymptotic  solutions of vortex equations are
available for both abelian Higgs and Chern-Simons Higgs gauge
theories. In this case   the analytic result  has been derived
for almost all of $n$ zero modes required by the index theorem
with the exception of only a finite number of modes with the
angular momentum $l/n\approx 1$ for the abelian Higgs and
$l/n\approx 0$ for the Chern-Simons case.  We have found that
the zero modes fill the core of the Abrikosov vortices but are
localized on the vortex boundary in  the Chern-Simons model.
The characteristic  thin-wall flux tube structure of the  giant
vortices in the  abelian Higgs model allows for the analytic
calculation of the zero modes even for the nonsupersymmetric
couplings. The general structure of the solution is  quite
robust and independent of fine details of the gauge and scalar
field interactions. Thus our analysis gives the answer to the
question posed in  Ref.~\cite{Witten:1984eb} about the
structure of the superconducting cosmic strings with $n>1$ for
the case of large $n$:  such strings  become  the thick
thin-wall flux tubes with the radius growing as $\sqrt{n}$  and
uniformly filled with the superconducting current density. For
the most phenomenologically interesting case of the
Jackiw-Rossi fermion coupling describing  the Majorana states
in topological superconductors, the  zero modes are formed
through a fine interplay between the effects of the magnetic
and scalar fields resulting in a set of the {warped} lowest
Landau level states. The remarkable profile of the resulting
density of states Eq.~(\ref{eq::rho}) can be used as a smoking
gun signature for experimental observation of the (effective)
Majorana fermions in two dimensions.

\vspace{2mm}
\noindent
{\bf Acknowledgement}\\[1mm]
The work of  L.G. was supported through by NSERC.
The work of  A.P. was supported in part
by NSERC and the Perimeter Institute for Theoretical Physics.

\end{document}